\RequirePackage{lineno}
\documentclass[twocolumn,epj,nopacs]{svjour}
\usepackage{graphicx}
\usepackage{textpos}
\usepackage{amssymb}
\usepackage{dcolumn}
\usepackage{bm}
\usepackage{multirow}
\usepackage{epsfig}
\usepackage{subfigure}
\usepackage{hyperref}

\newcommand{\snutau}{\ensuremath{\tilde{\nu}_{\tau}}}
\newcommand{\Memu}{\ensuremath{m_{e\mu}}}

\newcommand{\zp}{\ensuremath{Z'}}
\newcommand{\msnt}{\ensuremath{m_{\tilde{\nu}_{\tau}}}}
\newcommand{\mzp}{\ensuremath{m_{\zp}}}
\newcommand{\memu}{\ensuremath{m_{e\mu}}}
\newcommand{\pho}{\phantom{0}}
\newcommand{\ifb}{\mbox{fb$^{-1}$}}
\newcommand{\ipb}{\mbox{pb$^{-1}$}}
\newcommand{\TeV}{\mbox{TeV}}
\newcommand{\GeV}{\mbox{GeV}}
\newcommand{\pt}{\ensuremath{p_T}}
\newcommand{\et}{\ensuremath{E_T}}

\begin{document}
\title{Search for a heavy neutral particle decaying into an electron and a muon using 1~$\ifb$ of ATLAS data
\begin{textblock}{6}[0,1](0,-3){\small CERN-PH-EP-2011-140; Submitted to EPJC Letters.} \end{textblock}
}
\titlerunning{Search for a heavy particle decaying into an electron and a muon}
\author{The ATLAS Collaboration}
\authorrunning{The ATLAS Collaboration}
\institute{CERN, 1211 Geneva 23, Switzerland \email{atlas.publications@cern.ch}}
\date{\today}

\abstract{
A search is presented for a high mass neutral particle that decays directly to the $e^\pm \mu^\mp$ final 
state. The data sample was recorded by the ATLAS detector in $\sqrt{s}=7~\TeV$ $pp$ collisions at the LHC 
from March to June 2011 and corresponds to an integrated luminosity of 1.07~fb$^{-1}$. 
The data are found to be consistent with the Standard Model background. The high $e^\pm \mu^\mp$ mass 
region is used to set 95\% confidence level upper limits on the production of two possible new physics processes: 
tau sneutrinos in an $R$-parity violating supersymmetric model and $\zp$-like vector bosons in a lepton 
flavor violating model. 
}

\maketitle

Short-lived particles that decay into two oppositely signed leptons of different flavors, 
$e^{\pm} \mu^{\mp}$ ($e\mu$), $e^{\pm}\tau^{\mp}$ ($e\tau$), 
or $\mu^{\pm} \tau^{\mp}$ ($\mu\tau$), are predicted by a number of extensions to the 
Standard Model (SM). Examples include sneutrinos in $R$-parity violating (RPV) supersymmetric (SUSY) models~\cite{rpv_sneutrino}, 
and extra gauge $\zp$ bosons with lepton flavor violating (LFV) interactions~\cite{zprime_pap}. 
This Letter reports a search for an excess of high invariant mass $e\mu$ ($m_{e\mu}$) events over SM predictions in 
$pp$ collisions at $\sqrt{s} = 7\TeV$ at the LHC. The $e\mu$ final state is chosen due to its clean detector 
signature and low SM background in the high $\memu$ region. 
Similar searches with the $e\mu$ final state have been reported previously by the CDF, 
D0 and ATLAS Collaborations~\cite{RPVCDF1,RPVCDF2,RPVCDF3,RPVD01,RPVD02,ATLAS11}. 
In this Letter, we report an updated search with a data sample approximately 30 times larger than 
used for the previous ATLAS search~\cite{ATLAS11} with improved sensitivity to new physics.

The \mbox{ATLAS} detector~\cite{atlas_detector} is a multi-purpose particle 
physics apparatus with a forward-backward symmetric 
cylindrical geometry and near $4\pi$ coverage in solid angle\footnote{ATLAS uses a 
right-handed coordinate system with its origin at the nominal interaction point~(IP) 
in the centre of the detector and the $z$-axis along the beam pipe. The $x$-axis points 
from the IP to the centre of the LHC ring, and the $y$ axis points upward. Cylindrical 
coordinates $(R,\phi)$ are used in the transverse plane, $\phi$ being the azimuthal angle 
around the beam pipe. The pseudorapidity is defined in terms of the polar angle $\theta$ as $\eta=-\ln\tan(\theta/2)$.}.
The inner tracking detector~(ID) consists of a silicon pixel detector, a silicon 
microstrip detector, and a transition radiation tracker. The
ID is surrounded by a thin superconducting solenoid providing a
2~T magnetic field and by a hermetic calorimeter system, 
which provides three-dimensional reconstruction of particle showers up to $|\eta|<4.9$. 
For $|\eta|<2.5$, the electromagnetic calorimeter is finely
segmented and plays an important role in electron identification.
The muon spectrometer~(MS) is based on three large superconducting toroids arranged with an 
eight-fold azimuthal coil symmetry around the calorimeters. Three stations of drift tubes and 
cathode strip chambers enable precise muon track measurements, and resistive-plate and thin-gap 
chambers provide muon triggering capability.

\indent The data sample used in this analysis was collected using 
single lepton~($e$ or $\mu$) triggers, 
between March and June 2011. The  total integrated luminosity 
is $1.07 \pm 0.04$~$\ifb$~\cite{lumi}. 
The trigger efficiency is measured to be 100\%, with a precision of 1\%, 
for $e\mu$ candidates that pass the default selection criteria described below.

\label{sec:event_selections}
To select $e\mu$ candidates, the electron candidate is required to have
$\pt>25\GeV$ and to have pseudorapidity $|\eta|<1.37$ or $1.52<|\eta|<2.47$.
It is further required to pass the ``medium''~\cite{electronMedium} quality definition, 
which is based on the calorimeter shower shape, 
track quality, and track matching with the calorimeter cluster. 
In addition, the electron is required to be isolated in the calorimeter with
$\et^{\Delta R<0.4}<10\GeV$, where $\et^{\Delta R<0.4}$ is defined as the transverse energy
deposited in the calorimeter within a cone of radius $\Delta R=\sqrt{\Delta \eta^2 + \Delta \phi^2}=0.4$
around the electron cluster. 
Corrections have been applied to account for energy leakage from the electron and energy deposition 
inside the isolation cone due to additional $pp$ collisions.
The muon candidate must be reconstructed in both the ID and the MS, 
and have $\pt>25\GeV$ and $|\eta|<2.4$. 
Furthermore, the muon is required to be isolated in the ID with $\pt^{\Delta R<0.4}<10\GeV$, where
$\pt^{\Delta R<0.4}$ is defined as the scalar sum of the $\pt$ of tracks associated to the primary vertex,
within a cone of radius $\Delta R=0.4$ around the muon track. Only tracks with $\pt>1\GeV$ are used. 
Furthermore, only electrons separated from muons by $\Delta R>0.2$ are considered.

The $e\mu$ candidate events are required to have exactly one electron and one muon
with opposite charge satisfying the above selection criteria. 
Furthermore, events have to contain at least one primary vertex reconstructed
with at least three associated tracks of $\pt>500$ MeV.

The SM processes that can produce an $e\mu$ signature can be divided into two categories: processes such as 
$Z/\gamma^*$\\$\rightarrow \tau \tau$, $t\bar{t}$, single top, $WW$, $WZ$ and $ZZ$, which can produce electrons and 
muons in the final state, and processes, referred to as fake background in this Letter, 
such as $W/Z$+$\gamma$, $W/Z$+jets and multijet events where the photon 
or one or two jets are reconstructed as leptons. 

The contributions from processes listed in the first category as well as photon-related backgrounds are estimated using Monte Carlo~(MC) 
samples generated at $\sqrt{s}=7\TeV$. The detector response simulation~\cite{atlasMC} 
is based on the {\sc geant4} program~\cite{geant}. 
Lepton reconstruction and identification efficiencies, energy scales and resolutions in the MC 
are corrected to the corresponding values measured in the data in order to improve the modeling of the background.
The MC predictions are normalized to the data sample based on the integrated luminosity and 
cross sections of various physics processes. Top production is generated 
with {\sc mc@nlo}~\cite{mcnlo} for $t\bar{t}$ and single top, 
the Drell-Yan process is generated with {\sc pythia}~\cite{pythia}, and the diboson processes are generated with {\sc herwig}~\cite{herwig}. 
Higher order corrections have been applied to the cross sections predicted by these generators~\cite{FEWZ,Bonciani,MCFM}.
The $W/Z+\gamma$ contribution in the fake background comes from the $W(\rightarrow \mu\nu)\gamma$ and
$Z(\rightarrow \mu\mu)\gamma$ processes, where the photon is reconstructed as an electron. 
This background is estimated using events generated with {\sc madgraph}~\cite{madgraph}.

The uncertainties for the $t\bar{t}$ and single top cross sections are taken to be 
10\%~\cite{ttbarUncert} and 9\%~\cite{WtUncert}, respectively. The cross sections for 
$W/Z+\gamma$, $Z/\gamma^*\rightarrow \tau \tau$, $WW$, $WZ$ and $ZZ$ are assigned uncertainties of 10\%, 5\%, 7\%, 7\%, 
and 5\%, respectively; these uncertainties arise from the choice of PDF, from factorisation and renormalisation scale dependence and from $\alpha_s$ variations.
The integrated luminosity uncertainty and 
other smaller systematic uncertainties from the lepton trigger,
reconstruction and identification efficiencies, energy 
(momentum) scale and resolution have been added in quadrature and are included in the total uncertainty.

The remaining fake backgrounds arise from the $W/Z$+jets and multijet processes, 
where leptons are present from $b$- or $c$-hadron decays or at least one jet is misidentified as a lepton.
Such lepton candidates are collectively referred to as ``non-prompt leptons" in this Letter.
These jet fake backgrounds account for $\sim$30\% of the expected $e\mu$ data yield and are estimated from 
data using a $4 \times 4$ matrix background estimation method described below. 
A looser lepton quality selection (called loose lepton here) is defined for 
each lepton type in addition to the default quality 
selection (called tight lepton here). 
For loose muons, the isolation requirement is dropped. For loose electrons, the ``loose'' electron 
identification criteria as defined in Ref.~\cite{electronMedium} are used and the isolation requirement is also dropped.
The tight and loose lepton selections are then used to classify events where both leptons pass the loose requirements 
into four categories, depending on whether both leptons subsequently pass the tight requirement ($N_{pp}$), only one lepton 
fails the tight requirement and the other lepton passes the tight 
requirement ($N_{pf}$ or $N_{fp}$), or both leptons fail the tight requirement ($N_{ff}$). 
The sample composition can be estimated by solving a linear system 
of equations: $(N_{pp}, N_{pf}, N_{fp}, N_{ff})^T=\epsilon (N_{e\mu}, N_{e\mu^{\dagger}}, 
N_{e^{\dagger}\mu}, N_{e^{\dagger}\mu^{\dagger}})^T$, where 
$N_{e\mu}$ (or $N_{e^{\dagger}\mu^{\dagger}}$) is the number of events with 
two prompt leptons (or two non-prompt leptons), while
$N_{e\mu^{\dagger}}$ and $N_{e^{\dagger}\mu}$ are the numbers of events with 
one prompt lepton and one non-prompt lepton. The 
matrix $\epsilon$ contains the probabilities for a loose quality lepton to pass the tight quality selection 
for both prompt and non-prompt leptons. 
The probability for prompt leptons (non-prompt leptons) is estimated by applying the 
loose and tight selections on $Z/\gamma^* \rightarrow ee/\mu\mu$ events (a sample of dijet events).
To take into account the lepton $\pt$ dependence of the two probabilities, 
the matrix equation is inverted for each event, giving four weights, corresponding to the four combinations of prompt and non-prompt leptons. These weights are then summed over all events to yield 
the total number of events with one or more non-prompt leptons. 
The overall jet fake background is found to be $1175 \pm 32$ (stat) events.
The breakdown of these contributions is estimated to be $N_{e\mu^{\dagger}}=375 \pm 30$ (stat), 
$N_{e^{\dagger}\mu}=89 \pm 13$ (stat) and $N_{e^{\dagger}\mu^{\dagger}} = 711 \pm 8$ (stat).
The overall systematic uncertainty of 10\% comes mainly from the uncertainty on the probability for a loose quality 
non-prompt muon to pass the tight quality selection.

\indent Table~\ref{tab:data_mc_cutflow} shows the number of events selected
in data and the estimated background contributions with their uncertainties (both statistical and systematic uncertainties are included). 
A total of 4053 $e\mu$ candidates are observed, while
the expectation from SM processes is $4145 \pm 250$ events.
The $m_{e\mu}$ distribution is presented in Fig.~\ref{fig:data_MC_Mass} for data and background contributions. 
The distribution of observed events is compared to the expected background using a Kolmogorov-Smirnov test with statistical uncertainties only~\cite{KStest}. 
The test probability is 56\%, consistent with the absence of a new physics signal.

Table~\ref{tab:data_mc_cutflow-highmass} shows the numbers of observed and predicted background 
events in eleven high $e\mu$ mass regions. 
Good agreement is found for all mass regions and no statistically significant data excess is observed.
Limits are set on the contributions of new physics processes to the high mass region from two scenarios: 
the production of $\snutau$ in an RPV SUSY model and of an LFV $Z'$ in extra-gauge boson models.

\begin{table}
\begin{center}
\caption{\label{tab:data_mc_cutflow} Estimated backgrounds in the selected sample, together with the observed event yield. The total integrated luminosity is 1.07~$\ifb$.}
\begin{tabular}{cc}
\hline
\hline
  Process & Number of events \\ \hline
 $t\bar{t}$  & 1580 $\pm$ 170 \\
 Jet fake & 1175 $\pm$ 120 \\
 $Z/\gamma^* \rightarrow \tau \tau$  & 750 $\pm$ 60 \\
 $WW$          & 380 $\pm$ 31 \\
 Single top    & 154 $\pm$ 16 \\
 $W/Z+\gamma$ &\pho 82 $\pm$ 13 \\
 $WZ$          & 22.4 $\pm$ 2.3 \\
 $ZZ$          &\pho 2.48 $\pm$ 0.26 \\
\hline
 Total background  & 4145 $\pm$ 250 \\ \hline
 Data   & 4053 \\

 \hline
 \hline
\end{tabular}
\end{center}
\end{table}

\begin{table}
\begin{center}
\caption{\label{tab:data_mc_cutflow-highmass} Estimated total backgrounds in the selected sample, together with the observed event yields for 11 high $e\mu$ mass regions.}
\begin{tabular}{ccc}
\hline
\hline
  $\Memu$  & ~~Data & ~~~~~SM prediction \\  \hline
  $>200$ GeV  & 286  & 288 $\pm$ 22 \\
  $>250$ GeV  & 152  & 136 $\pm$ 11 \\
  $>300$ GeV  & 70  & 67 $\pm$ 6 \\
  $>350$ GeV  & 35  & 34.0 $\pm$ 3.0 \\
  $>400$ GeV  & 22  & 17.7 $\pm$ 1.7 \\
  $>450$ GeV  & 10  & 10.5 $\pm$ 1.2 \\
  $>500$ GeV  & 7  &\pho 6.8 $\pm$ 0.9 \\
  $>550$ GeV  & 3  &\pho 4.3 $\pm$ 0.6 \\
  $>600$ GeV  & 3  &\pho 2.4 $\pm$ 0.4 \\
  $>650$ GeV  & 1  &\pho 1.49 $\pm$ 0.31 \\
  $>700$ GeV  & 0  &\pho 1.07 $\pm$ 0.25 \\
 \hline
 \hline
\end{tabular}
\end{center}
\end{table}

The process $d\bar{d} \rightarrow \snutau \rightarrow e\mu$ in a SUSY RPV model is considered. 
The RPV sneutrino couplings allowed in the supersymmetric Lagrangian are 
$\frac{1}{2} \lambda_{ijk}\hat{L}_{i} \hat{L}_{j}
\hat{E}_{k} + \lambda^{'}_{ijk} \hat{L}_{i}
\hat{Q}_{j} \hat{D}_{k}$, where $L$ and $Q$ are the lepton and quark $SU(2)$ doublet superfields,
and $E$ and $D$ denote the singlet fields for charged leptons and down type quarks, respectively. The indices $i,j,k=1,2,3$
refer to the fermion generation numbers. The coupling constants $\lambda$ satisfy $\lambda_{ijk}=-\lambda_{jik}$. 
Only the tau sneutrino is considered in this Letter since stringent limits already exist on the electron sneutrino 
and muon sneutrino~\cite{rpv_sneutrino}. 
By fixing all RPV couplings except $\lambda^{'}_{311}$ ($\snutau$ to $d\bar{d}$) 
and $\lambda_{312}$ ($\snutau$ to $e\mu$) to zero,
and assuming that $\snutau$ is the lightest supersymmetric particle,
the contributions to the $e\mu$ final state originate from the $\snutau$ only.
The cross section is 0.154~pb for $\msnt=650\GeV$, $\lambda'_{311}=0.10$ and $\lambda_{312}=\lambda_{321}=0.05~$\cite{sneutrino_nlo}.
The total decay width is $\Gamma_{\snutau} = (3\lambda'^2_{311} + 2 \lambda^2_{312}) \msnt/16\pi$.
Using couplings that are consistent with the current limits,  
the decay width is less than $1\GeV$ for $\msnt=1\TeV$, 
which is well below the contribution from detector resolution.
MC samples with $\snutau$ masses ranging from 0.1 to $2\TeV$ are generated with {\sc herwig}~\cite{herwig,herwigSig}. 
 
An $e\mu$ resonance also appears in models containing a heavy neutral gauge boson, $Z'$~\cite{Langacker2}, 
with non-diagonal lepton flavor couplings. 
Rare muon decay searches have placed extremely stringent limits on the combination 
of the mass and the coupling to $ee$ and $e\mu$ in such models~\cite{zprime_pap}. The $e\mu$ searches at 
hadron colliders are not able to match the sensitivity of dedicated $\mu\rightarrow e$ 
conversion experiments. A limit on the production cross section times branching ratio to $e\mu$ 
is placed on the $\zp$-like boson model to represent the production of vector particles 
that can decay to the $e\mu$ final state.
To calculate the efficiency and acceptance, the $\zp$ is assumed 
to have the same quark and lepton couplings as the SM $Z$ except a non-zero $\zp$ to $e\mu$ coupling, 
which is assumed to be the same as the $\zp$ to $ee$ coupling.
The cross section is 0.61~pb for $\mzp=700\GeV$~\cite{zp_nnlo}.
MC samples with $\zp$ masses ranging from 0.7 to 2~TeV are generated with {\sc pythia}.

Both $\snutau$ and $\zp$ samples are processed through the standard chain of the \mbox{ATLAS} 
simulation and reconstruction. The overall product of acceptance and efficiency is 
36\% for $\msnt=100\GeV$ and increases to 64\% for $\msnt=1\TeV$. 
The corresponding number is~$\sim60\%$ for $\zp$ with mass $\mzp=700\GeV$ to $\mzp=2\TeV$.
The predicted $\memu$ distributions for a $\snutau$ with $\msnt=650\GeV$ and a $\zp$ with $\mzp=700\GeV$ are also
shown in Fig.~\ref{fig:data_MC_Mass}.

\begin{figure}[!ht]
\centering
\includegraphics[width=0.45\textwidth]{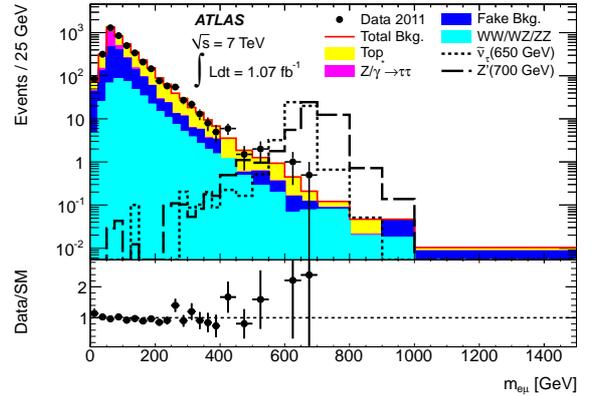}
\caption{Observed and predicted $e\mu$ invariant mass distributions.
Signal simulations are shown
for $\msnt=650\GeV$ and $\mzp=700\GeV$. The couplings $\lambda'_{311}=0.10$ and $\lambda_{312}=0.05$ 
are used for the RPV $\snutau$ model. 
The production cross section is assumed to be the 
current published limit of 0.178~pb for the LFV $\zp$ model~\cite{ATLAS11}. 
The ratio plot at the bottom includes only statistical uncertainties.}
\label{fig:data_MC_Mass}
\end{figure}

The $\memu$ spectrum is examined for the presence of a new heavy particle. 
For each assumed $\msnt$ value in the range 100~$\GeV$ to 2~$\TeV$, a search region, 
which depends on the simulated $e\mu$ mass resolution, is used\footnote{The search region is normally defined to be 
($\msnt-3\sigma$, $\msnt+3\sigma$), where $\sigma$ is the expected 
$\memu$ resolution (e.g., $\sigma=11$~GeV for $\msnt=400\GeV$). 
If $\msnt-3\sigma<700\GeV$ and $\msnt+3\sigma>700\GeV$, the region above 
$\msnt-3\sigma$ is used. If $\msnt-3\sigma>700\GeV$, the region above $700\GeV$ is used. 
The mass window changes around $700\GeV$ because the MC statistics is not sufficient in the $\memu>700\GeV$ region.}. 
The number of observed and predicted background and signal events in each search range are used to set an upper 
limit on $\sigma(pp \rightarrow \snutau) \times \mbox{BR}(\snutau \rightarrow e\mu)$.
A Bayesian method~\cite{Bayesian} is used with a uniform prior for the signal cross section for a given 
$\msnt$. Fig.~\ref{fig:results_parameter_limits}a shows the expected and observed 95\% confidence level (C.L.) limits, as a function of $\msnt$, 
together with the limits previously published by ATLAS~\cite{ATLAS11}, which were based on 35~$\ipb$ of data, and the expected $\pm 1$ and $\pm 2$ standard deviation uncertainty bands.
For a $\snutau$ with a mass of 100~GeV~(1~TeV), the limit on the cross section times branching ratio is 135~(4.5)~fb. 
The limits obtained extend 7~(34) times beyond the previous ATLAS results.
The theoretical cross sections for $\lambda'_{311}=0.10$, $\lambda_{312}=0.05$ and $\lambda'_{311}=0.11$,
$\lambda_{312}=0.07$ are also shown.
Tau sneutrinos with a mass below $1.32~(1.45)\TeV$ are excluded, assuming coupling values $\lambda'_{311}=0.10$ and $\lambda_{312}=0.05$ ($\lambda'_{311}=0.11$ and $\lambda_{312}=0.07$). 
The limits are significantly better than the limits from the previous ATLAS analysis using $35~\ipb$ of data.
The 95\% C.L. observed upper limits on $\lambda_{311}^{'}$ as a function of $\msnt$ are
shown in Fig.~\ref{fig:results_parameter_limits}b for three values of $\lambda_{312}$, together with the 
exclusion region obtained from the D0 experiment~\cite{RPVD02} and previously by the ATLAS experiment~\cite{ATLAS11}. 
The limits on $\lambda_{311}^{'}$ are tighter than the D0 results for $\msnt>270\GeV$ sneutrinos assuming $\lambda_{312}=0.07$.
Better sensitivity can be obtained for $\msnt<270$ GeV by applying selection cuts on missing transverse 
energy and number of jets in the event to improve the signal and background ratio, but it will make the 
search model-dependent.
\begin{figure}[htbp]
\centering
\includegraphics[width=0.45\textwidth]{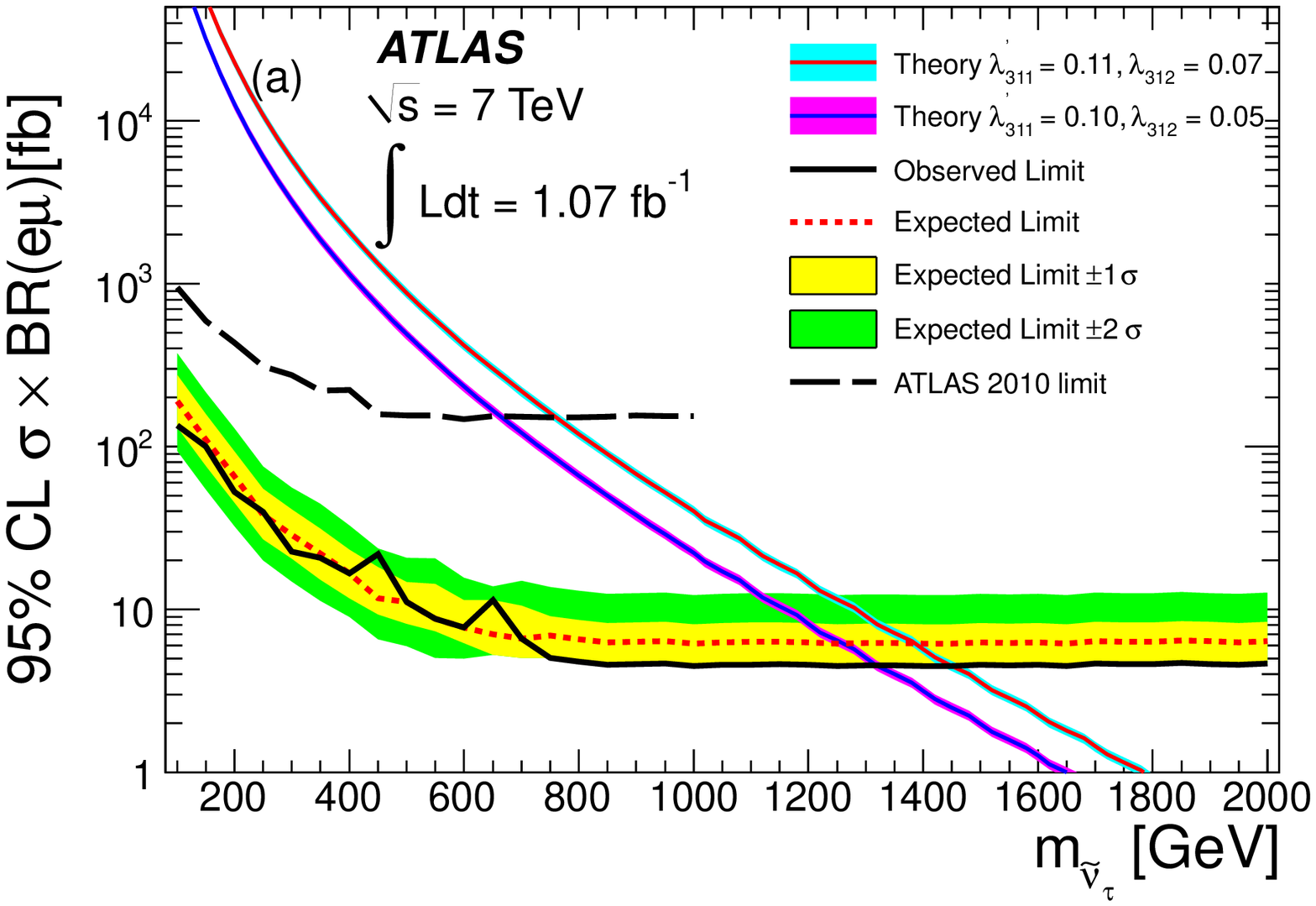}
\includegraphics[width=0.45\textwidth]{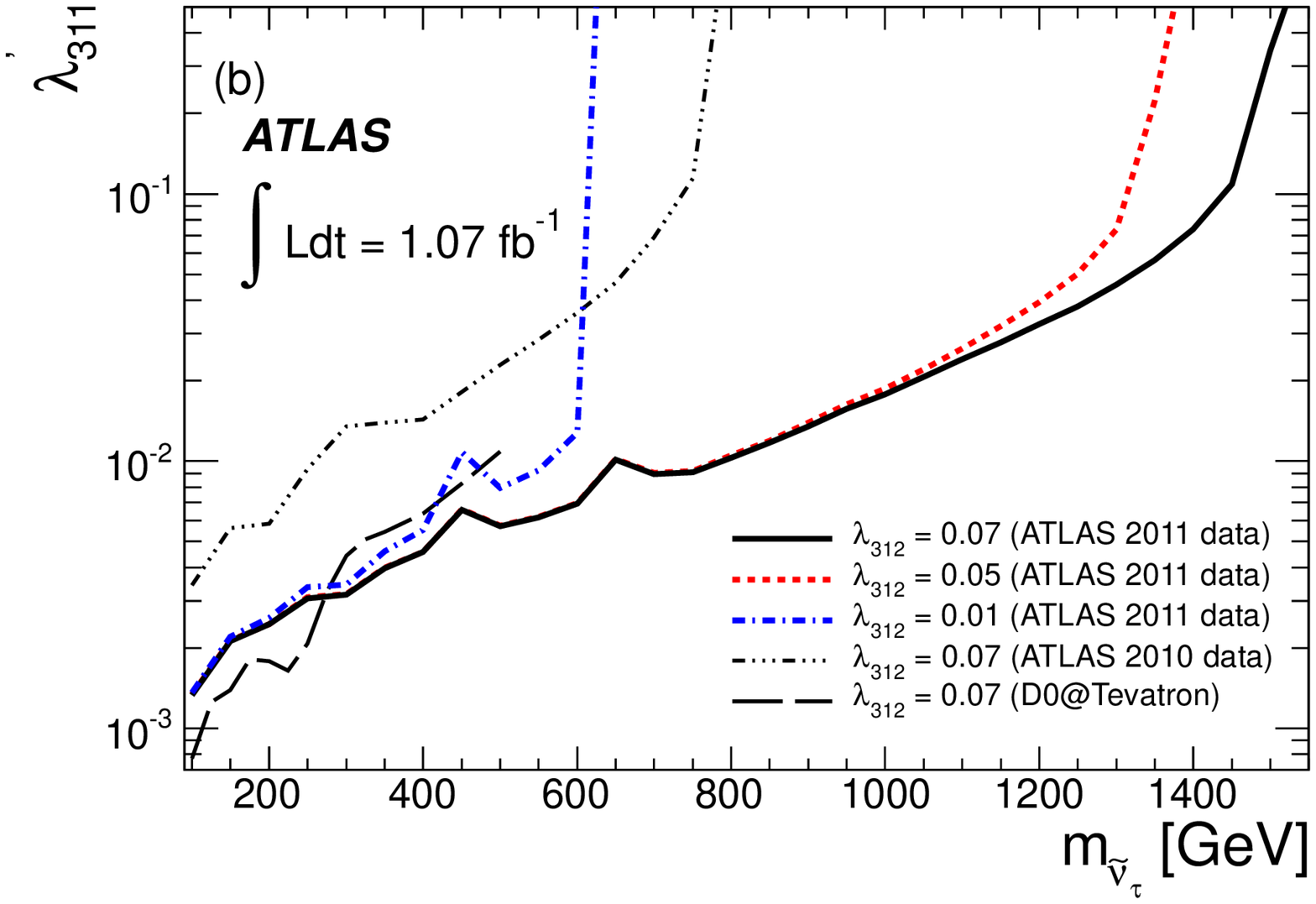}
\caption{ (a) The observed 95\% C.L. upper limits on
$\sigma(pp \rightarrow \snutau) \times \mbox{BR}(\snutau \rightarrow e\mu)$ as a function
of $\msnt$. The expected limits are also shown together with the expected $\pm 1$ and $\pm 2$ standard deviation
uncertainty bands. The previous ATLAS published limit and two theoretical cross sections 
for $\lambda'_{311}=0.10$, $\lambda_{312}=0.05$ 
and $\lambda'_{311}=0.11$, $\lambda_{312}=0.07$ 
calculated using {\sc madgraph} with next-to-leading order 
$k$-factors applied~\cite{sneutrino_nlo} are also shown.
(b) The 95\% C.L. upper limits on the $\lambda'_{311}$ coupling as a function of $\msnt$ for three values of $\lambda_{312}$. 
The regions above the three curves represent ranges of $\lambda'_{311}$ values that are excluded. 
These results are compared with the exclusion regions obtained from the D0 experiment and the previously 
published ATLAS analysis. The cross section times branching ratio for $pp \rightarrow e\mu$ is propotional 
to $\lambda^{'2}_{311} \lambda^{2}_{312}/(3\lambda^{'2}_{311} + 2\lambda^2_{312})$, which causes the 
weak dependence of the $\lambda^{'}_{311}$ limits on $\lambda_{312}$ for low mass tau sneutrinos.}
\label{fig:results_parameter_limits}
\end{figure}

A similar method is used to set limits on the LFV $\zp$-like vector boson; however, as opposed to the sneutrino limits, 
a unique mass window is defined for each potential signal mass. 
The 95\% C.L. upper limits on $\sigma(pp \rightarrow \zp) \times \mbox{BR}(\zp \rightarrow e\mu)$ are 
shown in Fig.~\ref{fig:results_Zprime_sigmaBR}. 
The expected limit is the same as the observed limit for the high mass points because 
both the median background event count expectation and the observed number of events are zero.
For a $\zp$ with mass of 0.7~TeV (1.0~TeV), the limit on the cross section times branching ratio is 9.6~fb~(4.8~fb). 
This result improves upon previous ATLAS limits by roughly a factor of 20~(40).

\begin{figure}[htbp]
\centering
\includegraphics[width=0.45\textwidth]{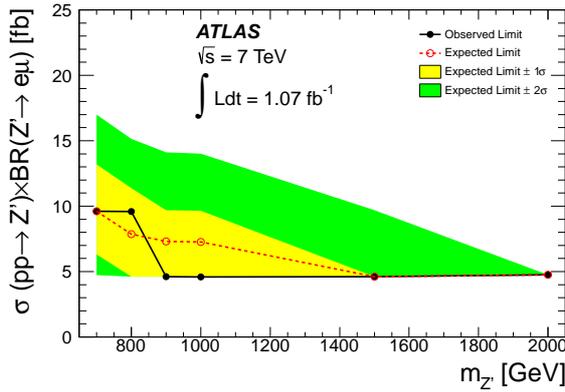}
\caption{\small The observed 95\% C.L. upper limits on $\sigma(pp \rightarrow \zp) \times \mbox{BR}(\zp \rightarrow e\mu)$.
The expected limits are also shown together with the expected $\pm 1$ and $\pm 2$ standard deviation
uncertainty bands. 
The observed and expected limits overlap as discussed in the text.
}
\label{fig:results_Zprime_sigmaBR}
\end{figure}

\indent In conclusion, a search has been performed for high mass $e\mu$ events using
$pp$ collision data at $\sqrt{s}=7\TeV$ recorded by the \mbox{ATLAS} detector. The observed 
$\memu$ distribution is found to be consistent with SM predictions.
With no evidence for new physics, 95\% C.L. exclusion limits are placed on the 
production cross sections and RPV coupling values of the tau sneutrinos in an RPV SUSY model, 
and tau sneutrinos with a mass below $1.32~(1.45)\TeV$ are excluded, 
assuming coupling values $\lambda'_{311}=0.10$ and $\lambda_{312}=0.05$ ($\lambda'_{311}=0.11$ 
and $\lambda_{312}=0.07$). The results presented here are the most stringent results to date for $\msnt>270$~GeV.
More stringent constraints are also set on the production cross sections of $\zp$ bosons in an LFV model. 
These two benchmark models can be used to represent the production of 
any narrow scalar and vector particles that can decay to the $e\mu$ final state. 

We thank CERN for the very successful operation of the LHC, as well as the
support staff from our institutions without whom ATLAS could not be
operated efficiently.

We acknowledge the support of ANPCyT, Argentina; YerPhI, Armenia; ARC,
Australia; BMWF, Austria; ANAS, Azerbaijan; SSTC, Belarus; CNPq and FAPESP,
Brazil; NSERC, NRC and CFI, Canada; CERN; CONICYT, Chile; CAS, MOST and
NSFC, China; COLCIENCIAS, Colombia; MSMT CR, MPO CR and VSC CR, Czech
Republic; DNRF, DNSRC and Lundbeck Foundation, Denmark; ARTEMIS, European
Union; IN2P3-CNRS, CEA-DSM/IRFU, France; GNAS, Georgia; BMBF, DFG, HGF, MPG
and AvH Foundation, Germany; GSRT, Greece; ISF, MINERVA, GIF, DIP and
Benoziyo Center, Israel; INFN, Italy; MEXT and JSPS, Japan; CNRST, Morocco;
FOM and NWO, Netherlands; RCN, Norway; MNiSW, Poland; GRICES and FCT,
Portugal; MERYS (MECTS), Romania; MES of Russia and ROSATOM, Russian
Federation; JINR; MSTD, Serbia; MSSR, Slovakia; ARRS and MVZT, Slovenia;
DST/NRF, South Africa; MICINN, Spain; SRC and Wallenberg Foundation,
Sweden; SER, SNSF and Cantons of Bern and Geneva, Switzerland; NSC, Taiwan;
TAEK, Turkey; STFC, the Royal Society and Leverhulme Trust, United Kingdom;
DOE and NSF, United States of America.

The crucial computing support from all WLCG partners is acknowledged
gratefully, in particular from CERN and the ATLAS Tier-1 facilities at
TRIUMF (Canada), NDGF (Denmark, Norway, Sweden), CC-IN2P3 (France),
KIT/GridKA (Germany), INFN-CNAF (Italy), NL-T1 (Netherlands), PIC (Spain),
ASGC (Taiwan), RAL (UK) and BNL (USA) and in the Tier-2 facilities
worldwide.

\clearpage
\onecolumn
\begin{flushleft}
{\Large The ATLAS Collaboration}

\bigskip

G.~Aad$^{\rm 48}$,
B.~Abbott$^{\rm 111}$,
J.~Abdallah$^{\rm 11}$,
A.A.~Abdelalim$^{\rm 49}$,
A.~Abdesselam$^{\rm 118}$,
O.~Abdinov$^{\rm 10}$,
B.~Abi$^{\rm 112}$,
M.~Abolins$^{\rm 88}$,
H.~Abramowicz$^{\rm 153}$,
H.~Abreu$^{\rm 115}$
% ...
\bigskip

% ...
$^{10}$ Institute of Physics, Azerbaijan Academy of Sciences, Baku, Azerbaijan\\
$^{11}$ Institut de F\'isica d'Altes Energies and Departament de F\'isica de la Universitat Aut\`onoma  de Barcelona and ICREA, Barcelona, Spain\\
% ...
$^{48}$ Fakult\"{a}t f\"{u}r Mathematik und Physik, Albert-Ludwigs-Universit\"{a}t, Freiburg i.Br., Germany\\
$^{49}$ Section de Physique, Universit\'e de Gen\`eve, Geneva, Switzerland\\
% ...
$^{88}$ Department of Physics and Astronomy, Michigan State University, East Lansing MI, United States of America\\
% ...
$^{111}$ Homer L. Dodge Department of Physics and Astronomy, University of Oklahoma, Norman OK, United States of America\\
$^{112}$ Department of Physics, Oklahoma State University, Stillwater OK, United States of America\\
% ...
$^{115}$ LAL, Univ. Paris-Sud and CNRS/IN2P3, Orsay, France\\
% ...
$^{118}$ Department of Physics, Oxford University, Oxford, United Kingdom\\
% ...
$^{153}$ Raymond and Beverly Sackler School of Physics and Astronomy, Tel Aviv University, Tel Aviv, Israel\\
% ...
$^{s}$ Also at Academia Sinica Grid Computing, Institute of Physics, Academia Sinica, Taipei, Taiwan\\
% ...
\end{flushleft}

\end{document}